\documentclass{article}

\usepackage[english]{babel}

\usepackage[letterpaper,top=2cm,bottom=2cm,left=3cm,right=3cm,marginparwidth=1.75cm]{geometry}

\usepackage{amsmath}
\usepackage{graphicx}
\usepackage[colorlinks=true, allcolors=blue]{hyperref}
\usepackage{amsthm}

\title{Quantifying deviations from Gaussianity with application to flight delays distributions}
\author{F. Olivares and M. Zanin \\
Instituto de F\'isica Interdisciplinar y Sistemas Complejos (CSIC–UIB), \\ Campus UIB, 07122 Palma, Spain.}

 \date{}
 
\begin{document}
\maketitle

\begin{abstract}
We propose a novel approach for quantifying deviations from Gaussianity by leveraging the permutation Jensen-Shannon distance. Using stable distributions as a flexible framework, we analyze the effects of skewness and heavy tails in synthetic sequences. We employ phase-randomized surrogates as Gaussian references to systematically evaluate the statistical distance between this reference and stable distributions. Our methodology is validated using real flight delay datasets from major airports in Europe and the United States, revealing significant deviations from Gaussianity, particularly at high-traffic airports. These results highlight systematic air traffic management strategies differences between the two geographic regions.
\end{abstract}

Ordinal Patterns; Jensen-Shannon Divergence; Air Traffic Management; Flight Delays; Non-Gaussian distributions; Stable Distributions 

\section{Introduction}

Several methodologies focus on estimating entropic quantifiers for characterizing the dynamical behavior of a system based on time series analysis~\cite{tang2015complexity}. In particular, structural entropies typically depend on a predefined probability distribution function, which transforms the time series from the time domain into the frequency/time-frequency domain or a symbolic representation. However, selecting an appropriate methodology---one capable of accurately capturing the intrinsic dynamical properties of the system---represents an ongoing challenge. This difficulty arises from the specific characteristics of the time series, such as its stationarity, length, and level of noise contamination~\cite{olivares2012ambiguities}. Notable techniques include, Fourier transformation~\cite{powell1979spectral}, wavelet transform~\cite{rosso2001wavelet}, procedures based on amplitude statistics~\cite{de2008randomizing}, binary symbolic dynamics~\cite{mischaikow1999construction}, and ordinal patterns~\cite{bandt2002permutation}.

Ordinal patterns provide a symbolic representation of 
a given time series by focusing solely on the relative ordering of its values~\cite{bandt2002permutation}. In this manner, the sequence of symbols emerges organically from the time series, eliminating the need for model-based assumptions.  Importantly, ordinal patterns account for the temporal information within the time series, assuming only a weak stationary condition~\cite{bandt2002permutation}. Thanks to these attributes, they have been widely applied to several scientific fields, including neuronal activity, epileptic events, self-similarity in nature, communications systems, sport science, finance, air transport, etc. (see Refs. \cite{amigo2023ordinal,zanin2012permutation,leyva202220} and references therein). 

One major drawback of using ordinal patterns is that they discard amplitude information~\cite{bandt2002permutation}. This omission can lead to loss of critical information, especially in systems where the absolute magnitude of fluctuations is important~\cite{cuesta2019permutation}, such as biomedical signals~\cite{fadlallah2013weighted}, and extreme event detection~\cite{azami2016amplitude}. To overcome this limitation, several improvements have been proposed~\cite{xiao2009fine,fadlallah2013weighted,azami2016amplitude,zanin2023continuous}.
A comparative study evaluating the classification performance of these approaches can be found in~\cite{cuesta2019permutation}. Nevertheless, all these approaches primarily focus on signal segmentation and spike detection~\cite{cuesta2019permutation}. Consequently, a quantitatively ordinal characterization of the corresponding amplitude distribution has not been thoroughly explored.   

In this study, we demonstrate that analyzing the ordinal pattern distribution of the walk of a time series, rather than its fluctuations, allows for a comprehensive characterization of its amplitude probability distribution. By leveraging the Permutation Jensen-Shannon Distance~\cite{zunino2022permutation}, a recently introduced metric that quantifies differences between two ordinal pattern distributions, we conduct a quantitative characterization of the ordinal dissimilarity between non-Gaussian and Gaussian distributions by treating phase-randomized surrogates of the former as reference sequences. We employ the family of stable distributions, which enable a quantitative analysis of heavy tails and asymmetry (skewness)~\cite{nolan2020univariate}. Notably, this distribution family extends the Central Limit Theorem to skewed heavy-tailed processes, providing a more accurate representation of real-world phenomena, such as commodity prices~\cite{mandelbrot1963new}. This property makes stable disributions particularly suitable for analyzing complex systems where Gaussian assumptions fail, including air traffic delays~\cite{mitsokapas2021statistical,wang2020universal,tu2008estimating,szabonon}. Understanding these deviations is crucial, as many forecasting and risk assessments models rely on normality assumptions~\cite{mueller2002analysis,wang2022distribution}. Consequently, detecting and quantifying these deviations is essential for improving predictive models~\cite{wang2022flight,hatipouglu2024predictive} and enhancing passenger satisfaction~\cite{song2020analyzing}. Here, our approach is validated using flight delay datasets from major airports in European and United States (US). The analysis reveals significant deviations from Gaussianity, particularly at high-traffic airports, with systematic differences between European and US air traffic management strategies.

\section{Ordinal patterns and the permutation Jensen-Shannon Distance}

The ordinal patterns probability distribution, introduced by Bandt and Pompe~\cite{bandt2002permutation}, relies on transforming a time series into a symbolic representation by encoding the relative ranking of the values within subsets of length $D$ ($D \in N$, the pattern length). Specifically, given a sequence of values $X(t) = \{x_t ; t = 1, \dots,M\}$, an ordinal pattern is defined by the permutation $\pi_{i}$ of the indices, $\{0, 1, . . . , D - 1\}$, that sorts the elements in ascending order. By counting the number of times each ordinal pattern $\pi_{i}$ appears in the encoded time series $X(t)$, normalised by the total number of ordinal patterns $M-(D-1)\tau$, the ordinal pattern probability distribution can be computed:
\begin{equation}\label{Eq:PDF} 
    p_{i} = \frac{ \# (\pi_{i})}{M-(D-1)\tau}, \,\,\, i = 1,2, ...,D!,
\end{equation}
where $\#(\pi_{i})$ stands for the cardinality of $\pi_{i}$. Note that the values of $X(t)$ can be mapped into subsets of length $D$ of consecutive ($\tau = 1$) or non-consecutive ($\tau > 1$) values. In their seminal work~\cite{bandt2002permutation}, Bandt and Pompe propose setting $\tau=1$ as a standard choice. However, selecting different values of $\tau$ can offer further insights into the characteristic scales of the system's dynamics, as it effectively represents multiples of the sampling time~\cite{zunino2010permutation,zunino2012distinguishing,olivares2020multiscale,olivares2023markov}. The ordinal pattern distribution has been widely employed in the computation of permutation entropy~\cite{bandt2002permutation}, which quantifies disorder based on the diversity of observed ordinal patterns~\cite{rosso2007distinguishing}.

Beyond entropy-based quantifiers, an important challenge in time series analysis is to measure the dissimilarity between different dynamical systems. To address this, the permutation Jensen-Shannon Distance~\cite{zunino2022permutation}, in short pJSD, has been introduced as a metric that quantifies the divergence~\cite{lin1991divergence} between two ordinal pattern distributions $P = \{p_1, ..., p_N\}$ and $Q = \{q_1, ..., q_N\}$ associated with two time series under analysis,
\begin{equation}\label{eq1}
    D_{\text{JS}}(P,Q) = S((P+Q)/2) - S(P)/2 - S(Q)/2,
\end{equation}
where $S(P)= \sum^{N}_{i=1} p_i \ln p_i $, is the classical Shannon entropy. The pJSD is hence obtained by calculating the square root of Eq. \ref{eq1}, and its normalised version reads
\begin{equation}\label{eq2}
    \text{pJSD}(P,Q) = \sqrt{\frac{D_{JS}(P,Q)}{\ln 2}},
\end{equation} 
which provides a bounded (pJSD $\in [0,1]$), symmetric, and interpretable measure of the ordinal distance between two time series. The pJSD provides a quantitative framework for comparing two time series by measuring how their ordinal pattern distributions differ. Furthermore, it has a key advantage, as hypothesis tests regarding the nature of an arbitrary time series can be easily performed by calculating its pJSD to reference time series generated in agreement with a null model~\cite{olivares2023markov} or surrogate analysis~\cite{zunino2024revisiting}. Note that, time series originating from the same underlying dynamical system are expected to exhibit small pJSD values, approaching asymptotically to zero following a power-law behavior with $M$ (pJSD $\propto M^{-1/2}$), but never exactly reaching it due to finite-size constraints~\cite{zunino2022permutation}.

\section{Numerical analysis}

\subsection{Stable Distributions}

\begin{figure*}[t!]
\includegraphics[width=\textwidth]{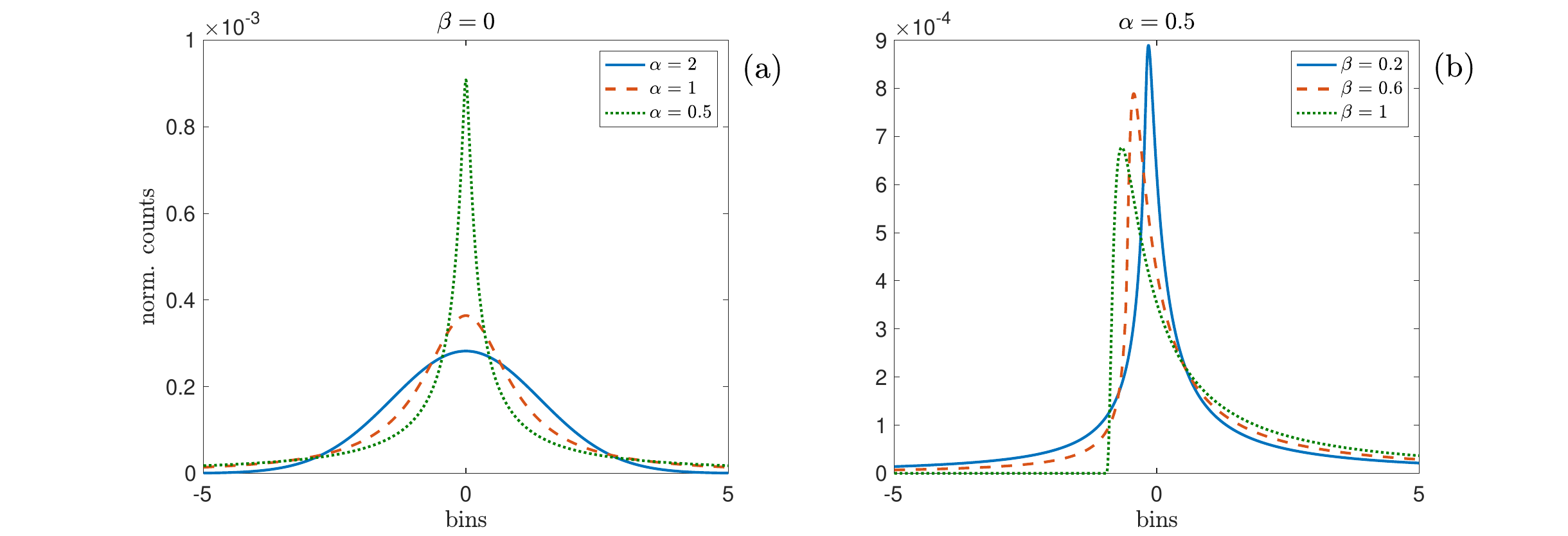}
\caption{\label{fig1} Probability density functions of stable distributions for the (a) symmetric cases with $\beta=0$ and $\alpha=2$ (Gaussian), $\alpha=1$ and $\alpha=0.5$, and (b) skewed cases with $\alpha=0.5$ and $\beta=0.2$, 0.6 and 1. In all cases,  $\delta=1$ and $\gamma=0$. }
\end{figure*}

A random variable $X$ is stable if and only if $X\stackrel{d}{=} \gamma Z + \delta$ (the symbol $\stackrel{d}{=}$ means equality in distribution), with $\gamma\neq 0$, $\delta \in R$, and $Z$ being a random variable with the characteristic function
\begin{equation}
    E \exp(i t Z) = \begin{cases}
    \exp \left[ -\left| t\right|^{\alpha} (1-i\beta \text{sign}(t) \tan \frac{\pi\alpha}{2})\right], &  \alpha \neq 1,\\[6pt]
    \exp \left[ -\left| t\right| (1+i\beta \frac{2}{\pi} \text{sign}(t)\log\left| t\right|\right] , &  \alpha = 1,
    \end{cases}
\end{equation}
with $t \in R$. The above definition shows that a family of Stable Distributions (SD) is defined by four parameters: a stability exponent $\alpha \in (0 , 2]$, a skewness parameter $\beta \in [-1,1]$, a scale parameter $\delta \in (0, \infty)$, and a location parameter $\gamma \in (-\infty, \infty)$~\cite{nolan2020univariate}. It is important to note that the scale parameter is not the same as the standard deviation (even for Gaussian case $\alpha=2$), and the location parameter is not typically the mean. The parameters $\alpha$ and $\beta$ determine the shape of the distribution. On one hand, as $\alpha$ grows, the weights of the tails of the distribution decrease until converging to Gaussiniality for $\alpha =2$, as shown in Fig. \ref{fig1}(a) with $\beta=0$, i.e. symmetric case. On the other hand, asymmetry is determined by $\beta$, as it increases/decreasing from zero, the distribution skews to the right/left. This feature is depicted in Fig. \ref{fig1}(b) with $\alpha=0.5$ and $\beta>0$. It is important to note that as $\alpha$ increases, the effect of $\beta$ decreases, and for $\alpha=2$, the skewness of the distribution in independent of $\beta$. 

\subsection{Synthetic data generation}

We consider synthetic uncorrelated time series sampled from this family of stable distributions for different values of $\alpha \in [0.5,2]$ with a step of $\Delta \alpha = 0.1$ and skewed distributions $\beta \in [-1,1]$ with a step of $\Delta \beta = 0.1$. For generating the synthetic sequences $x(t)$, we have used the MATLAB function \texttt{random} that follows the parametrization described in Ref.~\cite{nolan2020univariate}. We have generated one hundred independent realizations of length $M=10^{5}$ data points. Then, the walk of the synthetic data is calculated as $Y_i =\sum_{t=1}^{i} (x_t-\gamma)$ with $i \in \{1, 2,3, ..., M\}$. Note that the location parameter $\gamma$ is equal to the mean of the distribution for $\alpha > 1$. For the present analysis we set $\gamma=0$ and $\delta=1$.

\subsection{Surrogate analysis}
It is our aim here to characterize deviation from Gaussianity as the shape parameters $\alpha$ and $\beta$ change, by measuring the distance between the ordinal distribution of the synthetic data to the one obtained from the phase-randomized surrogate realizations~\cite{theiler1992testing}, which establishes a Gaussian reference. Nevertheless, these Gaussian references can still yield positive values of the pJSD due to statistical fluctuations. For tackling this, the distances are normalised by defying a Z-Score using the average, $\langle pJSD_{FFT}\rangle$, and the standard deviation, $\sigma_{pJSD_{FFT}}$ of the values resulting from the distance between a pair of two phase-randomized realizations from the synthetic data:
\begin{equation}
    Z_{pJSD} = \frac{pJSD-\langle pJSD_{FFT}\rangle}{\sigma_{pJSD_{FFT}}}.
\end{equation}
One hundred independent realizations have been considered. As a result, the normalised distance, $Z_{pJSD}$, quantifies how many standard deviations a given stable distribution is away from the Gaussian reference dataset. The normalised distances are expected to be zero for distributions that coincide with the average obtained by the phase-randomized surrogate set. Conversely, positive values indicate the magnitude (in standard deviations) of the deviation from that average. Normally, for $Z_{pJSD} < 3$ ($99.7 \% $), the observed distribution is considered Gaussian. We set $\tau=1$, since temporal information is not relevant for our analysis.

\begin{figure*}[h!]
\includegraphics[width=\textwidth]{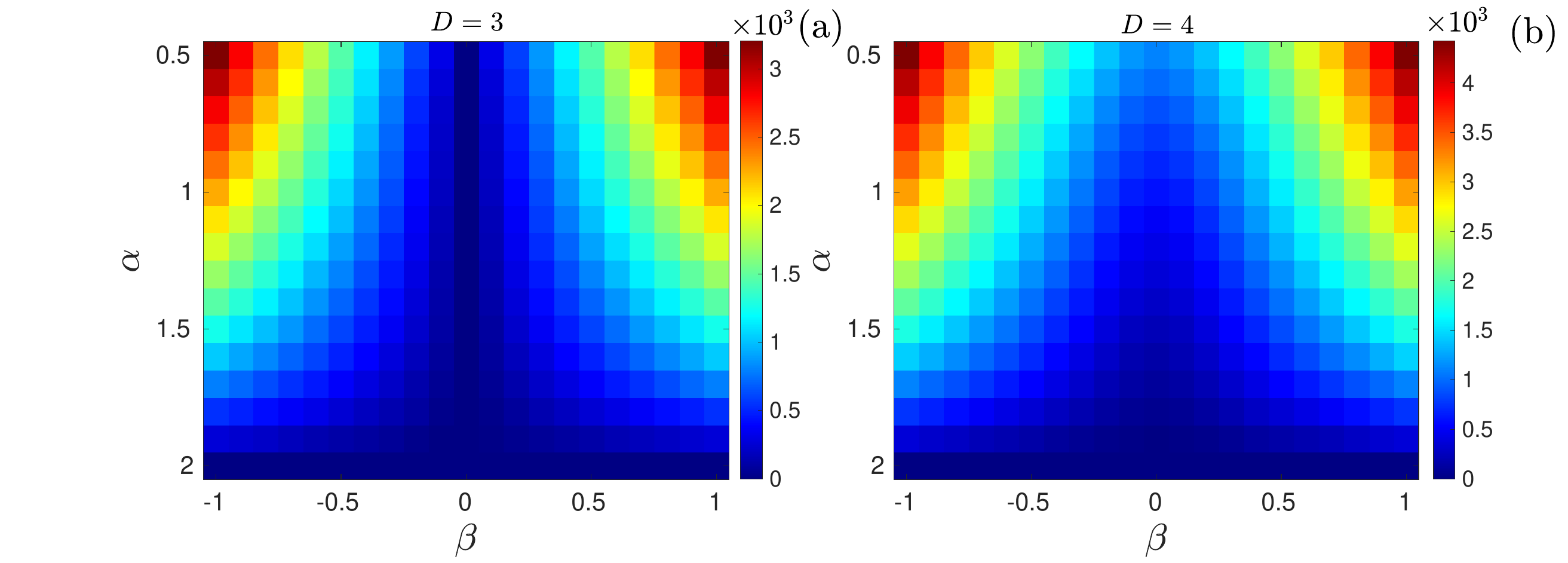}
\caption{\label{fig2} $Z_{pJSD}$ between sequences following Stable Distribution and their phase randomized surrogates as a function of the parameter $\beta\in [-1,1]$ and $\alpha \in [0.5,2]$ with a step of 0.1, for (a) $D=3$ and (b) $D=4$. Synthetic sequences of $M=10^{5}$ data points. Average over 100 independent realizations is reported.  }
\end{figure*}

The $Z_{pJSD}$ as a function of $\beta$ and $\alpha$ is shown in Fig. \ref{fig2} for $D=3$ (a) and $D=4$ (b). For the symmetric case, $\beta=0$, the pattern length $D=3$ does not detect deviation from Gaussianity as $\alpha$ decreases, i.e. the presence of heavy tails. This is not the case for $D=4$. For $\beta \in [-1,0) \cup (0,1]$, the $Z_{pJSD}$ correctly quantifies deviation from Gaussianity indistinctly of the sign of the skewness. Naturally, the distance depends on the value of the pattern length since the larger the value of $D$ the more information about the distribution is captured. Finally, we found that for $\alpha=2$ the distance remains close to zero independently of the pattern length.

\begin{figure*}[ht!]
\includegraphics[width=\textwidth]{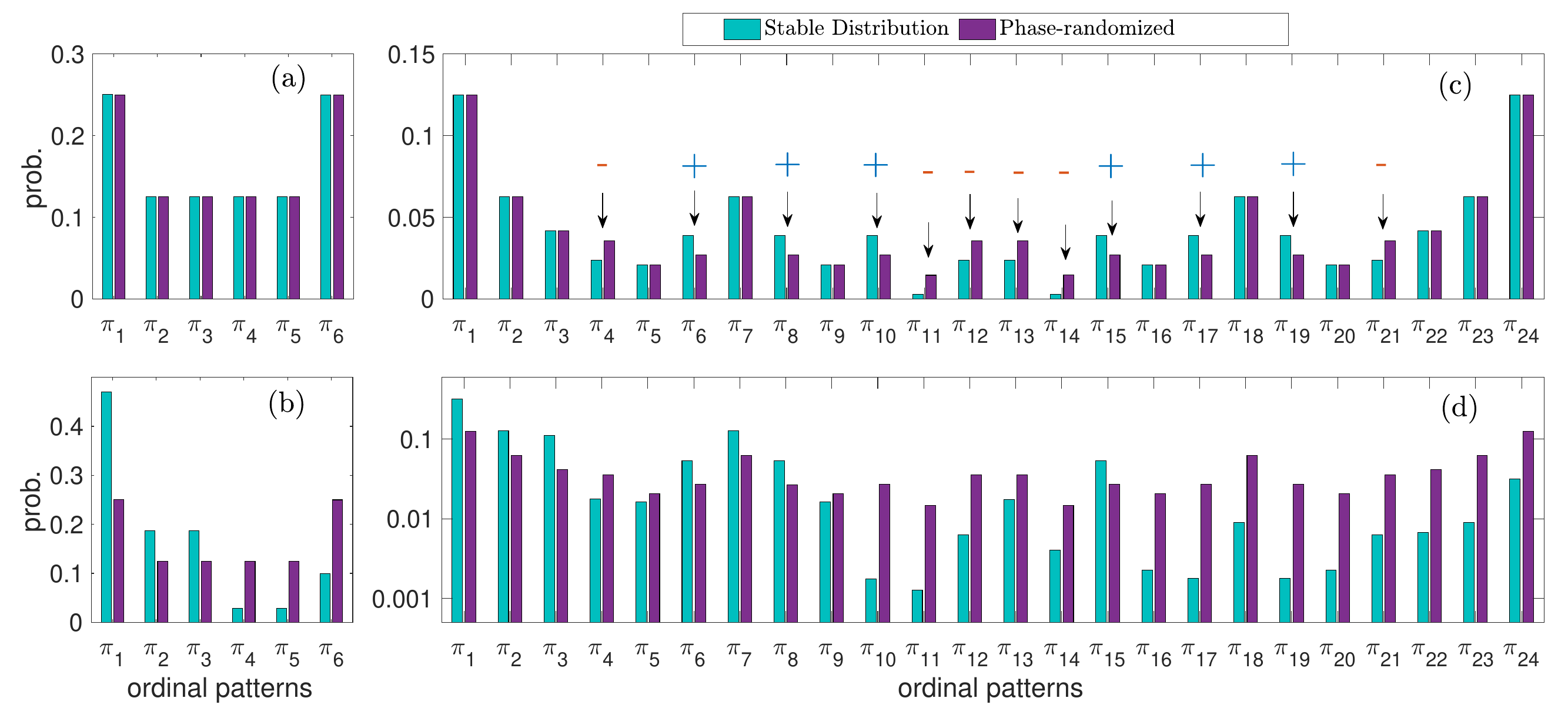}
\caption{\label{fig3} Ordinal pattern probabilities obtained from a sample data following a Stable Distribution with $\alpha=0.5$ and (a) $\beta=0$ and $D=3$, (b) $\beta=1$ and $D=3$, (c) $\beta=0$ and $D=4$ and (d) $\beta=1$ and $D=4$. Average over 100 independent realizations is reported. Arrows indicate the probabilities that mismatch their phase-randomized counterpart, with the sign symbols representing if they are larger or lower than them.  }
\end{figure*}

It is well known that Gaussian processes have a symmetric ordinal pattern distribution as a results of the distribution of the original amplitude values~\cite{bandt2007order}. For totally uncorrelated data and $D=3$, the probabilities are clustered into two groups: $p_{\pi_1}=p_{\pi_2}=1/4$ and $p_{\pi_2}=p_{\pi_3}=p_{\pi_4}=p_{\pi_5}=1/8$. We found that such a symmetric clustering is still valid for $\alpha<2$ and $\beta=0$, as can be observed in Fig. \ref{fig3}(a), that depicts the case of $\alpha=0.5$ as a representative result of all values of $\alpha \in [0.5,2]$. The distribution matches the one obtained from the phase-randomized data (Gaussian). Same results are found for other values of the stability exponent and $\beta=0$. This inherited property is the reason why the deviation from Gaussianity when $\beta=0$ is not captured by the $Z_{pJSD}$ for $D=3$---see Fig. \ref{fig2}(a). On the contrary, for a positive skewed distribution, the symmetric clustering property is not longer valid as observed in Fig. \ref{fig3}(b). This result is qualitatively representative for other values of $\beta$. For $D=4$, we found quite interesting results for the symmetric case. The ordinal patterns distribution is still symmetrically clustered as its Gaussian counterpart, yet, some probability weights differ, as observed in Fig. \ref{fig3}(c). The probabilities that differ are marked with an arrow. The mismatch weights are the responsible for the $Z_{pJSD}$ detecting the symmetric deviation from Gaussianity---see Fig. \ref{fig2}(b). For the sake of comparison, Fig. \ref{fig3}(d) shows the ordinal distribution for the asymmetric case $\beta=1$ (note that the $y$ axis is in logarithmic scale for better visualization of the probabilities). Same conclusions, as for $D=3$, can be drawn from this result. For negative skewed cases, $\beta<0$, the ordinal pattern distribution is a mirror image of the one obtained for the positive skewed cases. Similar results are found for $D=5$ and 6.

\begin{figure*}[h!]
\includegraphics[width=\textwidth]{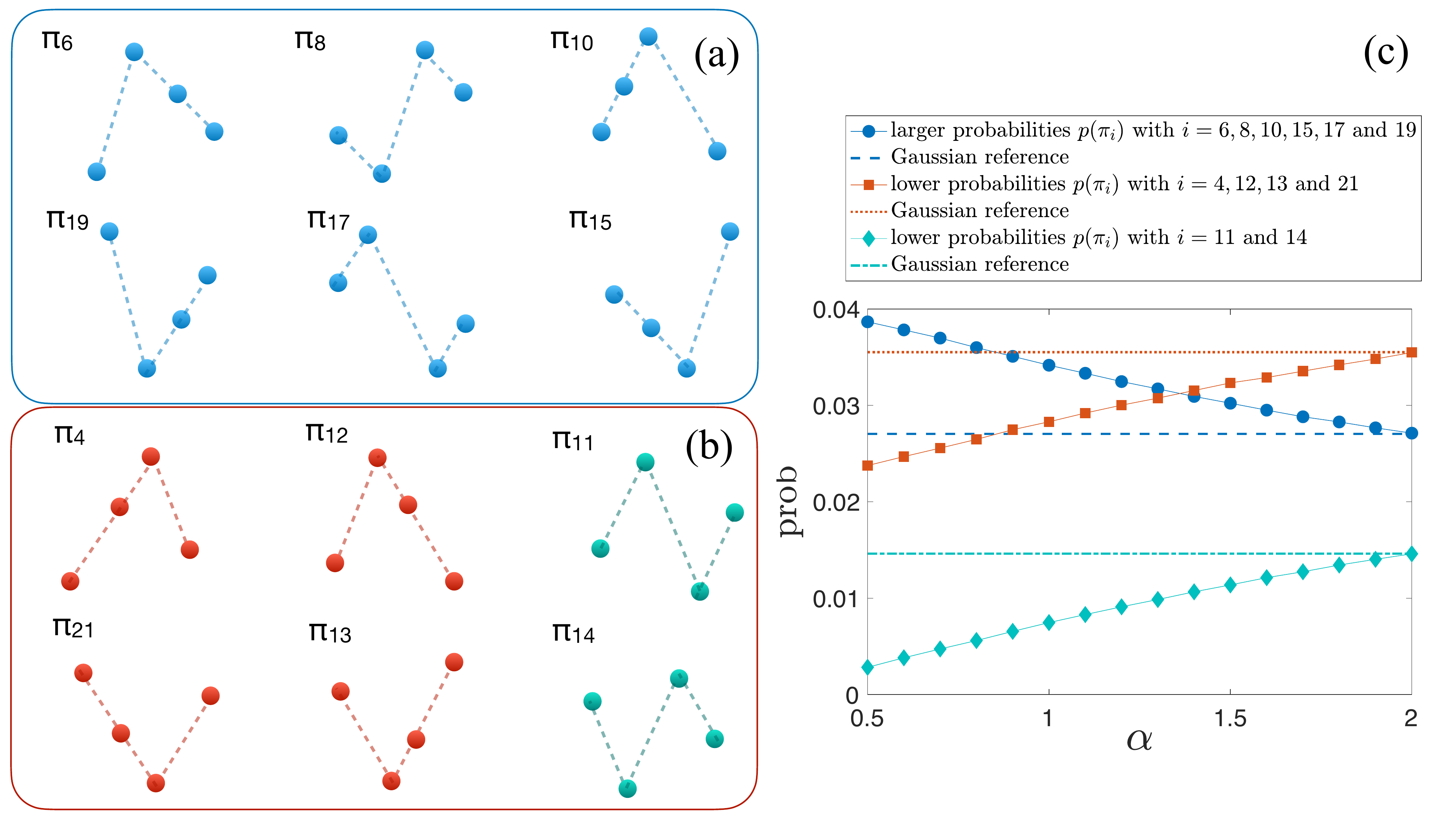}
\caption{\label{fig4} Ordinal patterns whose probability of occurrence is (a) larger and (b) lower than the ones obtained from phase-randomized data. (c) Ordinal pattern probabilities for $D=4$ that differ from their phase-randomized counterpart for the symmetric case ($\beta=0$) as a function of $\alpha$ }
\end{figure*}

Let us on the symmetric case. From Fig. \ref{fig3}(c) we observe that for $D=4$ some patterns are less frequent ($\pi_{4}$, $\pi_{11}$, $\pi_{12}$, $\pi_{13}$, $\pi_{14}$ and $\pi_{21}$), while other are more frequent ( $\pi_{6}$, $\pi_{8}$, $\pi_{10}$, $\pi_{15}$, $\pi_{17}$ and $\pi_{19}$) than their phase-randomized counterparts---see arrows in Fig. \ref{fig3}(c). These patterns are represented in Fig. \ref{fig4}(a) and (b) respectively. As the stability exponent $\alpha$ decreases (departures from Gaussianity) the tails of the distribution become heavier, \emph{i.e.} the probability of large values becomes higher, which is the trigger of having more jumps in the walk. It can straightforwardly be concluded that these jumps are translated to the ordinal patterns depicted in Fig. \ref{fig4}(a). While the occurrence of these patterns increases with $\alpha$, other permutations' frequencies decrease which are represented in Fig. \ref{fig4}(b). The evolution of all these probabilities with $\alpha$ is shown in Fig. \ref{fig4}(c). We observe that they gather into 3 groups. For $\alpha=2$, all the probabilities match the Gaussian reference---dotted, dashed and dotted-dashed lines. As $\alpha$ decreases the 6 jump-related patterns increase their probability, while the remaining patterns do the opposite. It becomes evident that ordinal patterns decode information about the amplitude when analyzing the walk of non-Gaussian increments.  

\begin{figure*}[h!]
\begin{center}
\includegraphics[width=0.7\textwidth]{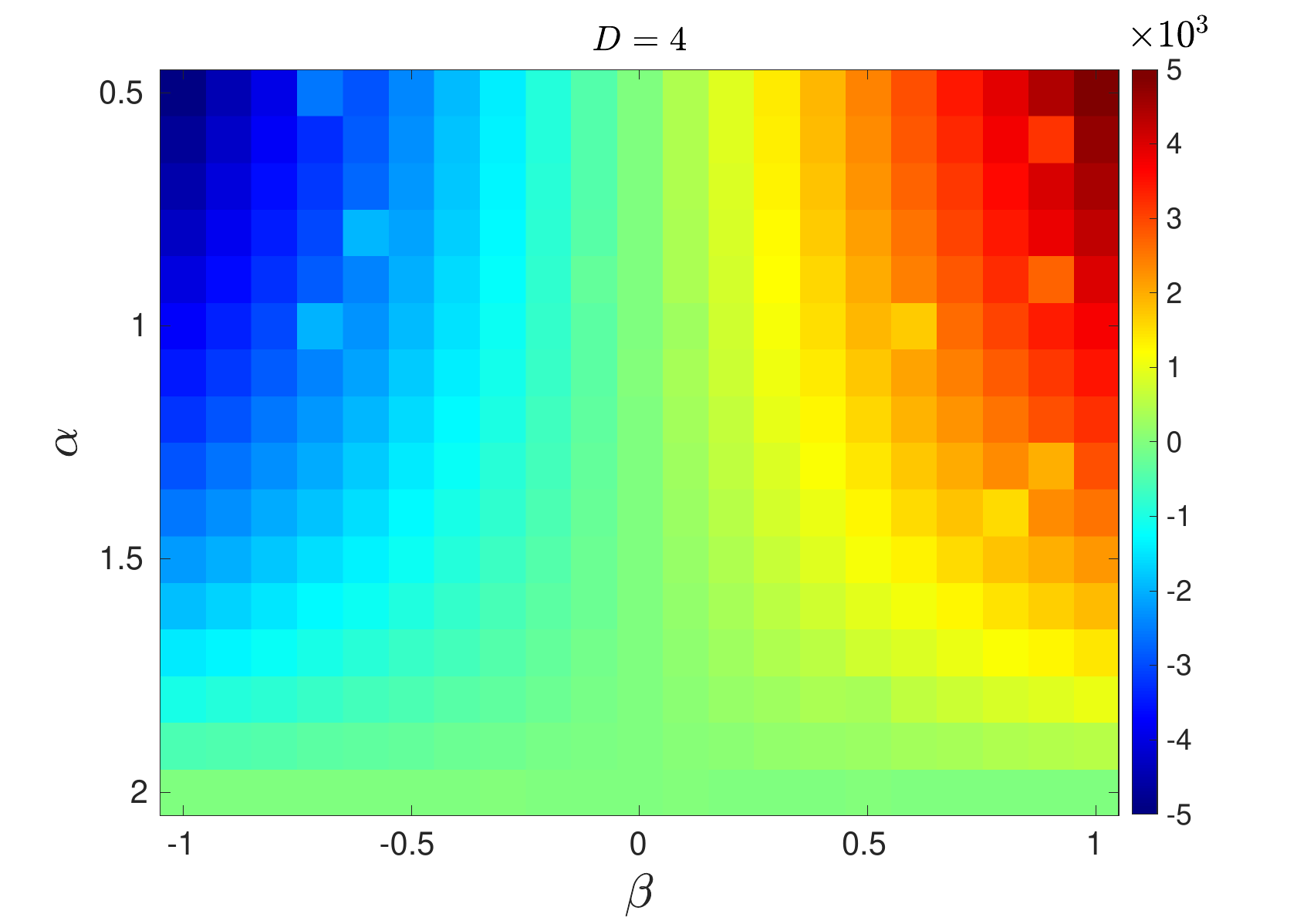}
\end{center}
\caption{\label{fig5} $Z_{pA}$ of the ordinal pattern distribution of the walk from sequences following Stable Distribution as a function of the parameter $\beta$ and $\alpha$. Synthetic sequences of $M=10^{5}$ data points. Average over 100 independent realizations is reported. }
\end{figure*}

In order to further quantify asymmetric deviations from Gaussianity ($\beta \neq 0$), we define a permutation asymmetry metric by computing the average of the sum of differences between symmetrically paired terms as
\begin{equation}
    \text{pA}(D) = \frac{2}{D!} \sum^{D!/2}_{i=1} \left(p_i - p_{D!/2 - (i-1)} \right).
\end{equation}
Here, $p_i$ represents the $i$-th term of the distribution, and $p_{D!/2 - (i-1)}$ represents the symmetrically opposite term in the distribution. In this manner, we can quantify the sign of the amplitude asymmetry of the data, \emph{i.e.} the sign of $\beta$. Additionally, for normalising this ordinal asymmetry metric we use the same approach applied to the pJSD, defining a Z-Score using the average and the standard deviation of the values resulting from the asymmetry of a set of one hundred independent phase-randomized realizations from the synthetic data:
\begin{equation}
    Z_{pA} = \frac{pA-\langle pA_{FFT}\rangle}{\sigma_{pA_{FFT}}}.
\end{equation}
Fig. \ref{fig5} depicts the results of this analysis for the data following a Stable Distribution. We observe that as the value of $\beta$ moves away from zero, the asymmetry $Z_{pA}$ correctly quantifies the sign of the skewness. Moreover, for $\beta=0$, the permutation asymmetry is equal to zero independently of the value on $\alpha$. Naturally, for $\alpha=2$, the $Z_{pA}$ is also zero. In this manner, the  $Z_{pJSD}$  and the  $Z_{pA}$ characterize the entire distribution of the increments of the walk.


\section{Application to flight delays distributions}

The appearance of delays in air transport is both a common occurrence and one of its main problems. Beyond negatively impacting the perception that customers have of this transportation mode, delays have profound implications in the cost-efficiency~\cite{cook2011european} and safety of the system~\cite{duytschaever1993development}; they further negatively impact the environment by unnecessarily increasing carbon dioxide emissions~\cite{carlier2007environmental}. Broadly speaking, these can be caused by random occurrences, as e.g. adverse weather phenomena or equipment failures; but also from systemic inefficiencies. We are here going to analyze a large data set of real operations, both in Europe and US; and show how the proposed metric of distance from Gaussianity can help to discriminate between the two sources.

\subsection{Analysis of delays' distributions}

As a consequence of the importance of delays in air transport, many studies have tried to provide a deeper and more comprehensive understanding of the factors and dynamics behind flight delays and their propagation, also focusing on their distributions. Note that, were delays of a completely stochastic nature, such statistical approach would indeed be motivated.

Among the first works to use this approach, Mueller et al.~\cite{mueller2002analysis} provided a comprehensive characterization of delay distributions at ten major US airports over a 21-day period. They found that departure delays are best modeled using a Poisson distribution, while en-route and arrival delays follow a Normal distribution.
Subsequently, Tu et al.~\cite{tu2008estimating} found that the departure delay distribution was best modeled using a finite mixture of four normal distributions, exhibiting heavy tails and right skewness, reflecting the different operational and stochastic factors affecting flight delays. The model was trained and validated using United Airlines flight data from Denver International Airport for the years 2000–2001.
More recently, Z. Szab\'o found that the delay distribution in both Europe and US are well fitted by a non-central Student's t-distribution~\cite{szabonon}. 

Moving to the analysis of causes, Cao et al.~\cite{cao2019method} classified all influencing factors into propagation and non-propagation factors (NPF). To illustrate, bad weather can randomly delay a few operations (what are also called ``primary'' delays); conversely, those delays may spread to subsequent operations of the same aircraft (i.e. generating ``secondary'' or ``reactionary'' delays). Considering the departure delays for Delta Airline, they showed that the ones attributed solely to NPFs adhere to a power law distribution, while the overall flight departure delay distribution follows a shifted power law. On the other hand, Wang and co-workers~\cite{wang2020universal} extended the analysis to multiple operators, and divided departure delays from 14 American airlines into the two aforementioned groups; they respectively exhibited a shifted power law or an exponentially truncated shifted power law decay. To further investigate delay distributions, by a comprehensive statistical analysis of flight delays at major UK airports, Mitsokapas et al.~\cite{mitsokapas2021statistical} identified a power-law decay in large positive delays and an exponential decay in early arrivals. 

On this topic, it is further worth considering the work of Wang and coauthors~\cite{wang2022distribution}, proposing machine learning models to predict the distributions of flight delays, and validating the methods using empirical data from Guangzhou Baiyun International Airport. They examined multiple probability distributions for modeling flight delays, including Beta, Erlang, and Normal distributions. The results suggested that Normal distribution is better able to capture the stochastic nature of flight delay.

\subsection{Real operational data}

We extracted time series of delays for aircraft arriving at the 50 largest European and US airports, ranked according to the respective number of operations. Data for Europe have been obtained from the EUROCONTROL’s R$\&$D Data Archive, freely accessible at \url{https://www.eurocontrol.int/dashboard/rnd-data-archive}, and corresponding to all operations executed throughout March, June, September and December between years 2015 and 2019. Note that these four months correspond to a limitation at source, and does not correspond to a decision of the authors.
Data for US have been obtained from the Reporting Carrier On-Time Performance database of the Bureau of Transportation Statistics, U.S. Department of Transportation, freely accessible at \url{https://www.transtats.bts.gov}. In order to obtain comparable results, data have been filtered according to the same dates as in the European set.

\begin{figure*}[t!]
\includegraphics[width=0.8\textwidth]{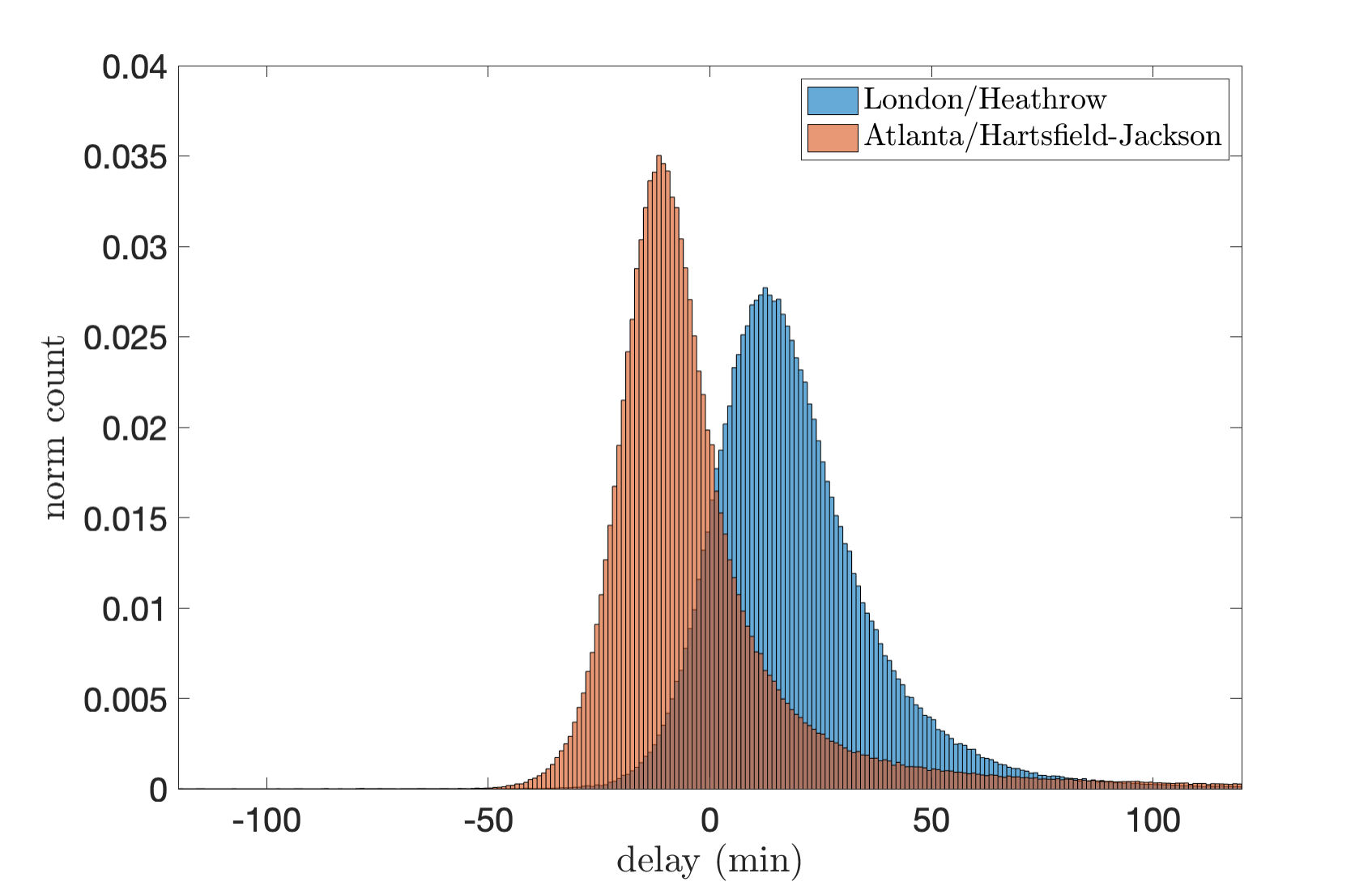}
\caption{\label{fig6} Distribution of the arrival delays in minutes for London/Heathrow and Atlanta/Hartsfield-Jackson, i.e. the biggest airports in Europe and USA, respectively.}
\end{figure*}

In both cases, two delays have been calculated for each flight: the arrival (or landing) delay, defined as the difference between the actual and planned landing times; and the en-route delay, calculated as the increase in delay observed between the take off and the landing. In order to create two individual time series per airport (i.e. one for arrival and one for en-route delays), these values have been concatenated according to the arrival sequence, in which each element is the delay $\delta_t$ of a single aircraft. To eliminate any daily and weekly oscillations~\cite{olivares2022corrupted,olivares2023measuring} between aircraft, we have shuffled all sequences. In that manner, the  distribution of ordinal patterns will only capture the information of the shape of the amplitude distribution. For illustration, Fig. \ref{fig6} shows the distribution of arrival delays for London-Heathrow and Atlanta airports, the biggest airports respectively in Europe and US.  It can be appreciated that both distributions are asymmetric and positive skewed around their mode. 

Following the previous methodology, we here compute the walk of the individuals delays $\delta_t$ as $\mathcal{D}_i =\sum_{t=1}^{i} (\delta_t - \gamma)$ with $i \in \{1, 2,3, ..., N_v\}$ and $N_v$ the number of arrivals of each airport. Normally, the parameter $\gamma$ would represent the mean of the distribution. However, as we aim to characterize the interplay between early (negative delays) and late arrivals (positive delays), the mean does not adequately represent the appropriate reference value; zero, i.e. the operation according to the plan, instead does. Similar to when we want to determine whether a distribution of financial returns is symmetric with respect to zero (i.e., if there are more days with large losses or large gains), we can define a skewness around zero as the reference point. Thus, by setting $\gamma=0$, we will be quantifying the exchange between severe delays and early arrivals. For all analysis below, we set $D=4$ and $\tau=1$.

\subsection{Results}
\begin{figure*}[h!]
\includegraphics[width=\textwidth]{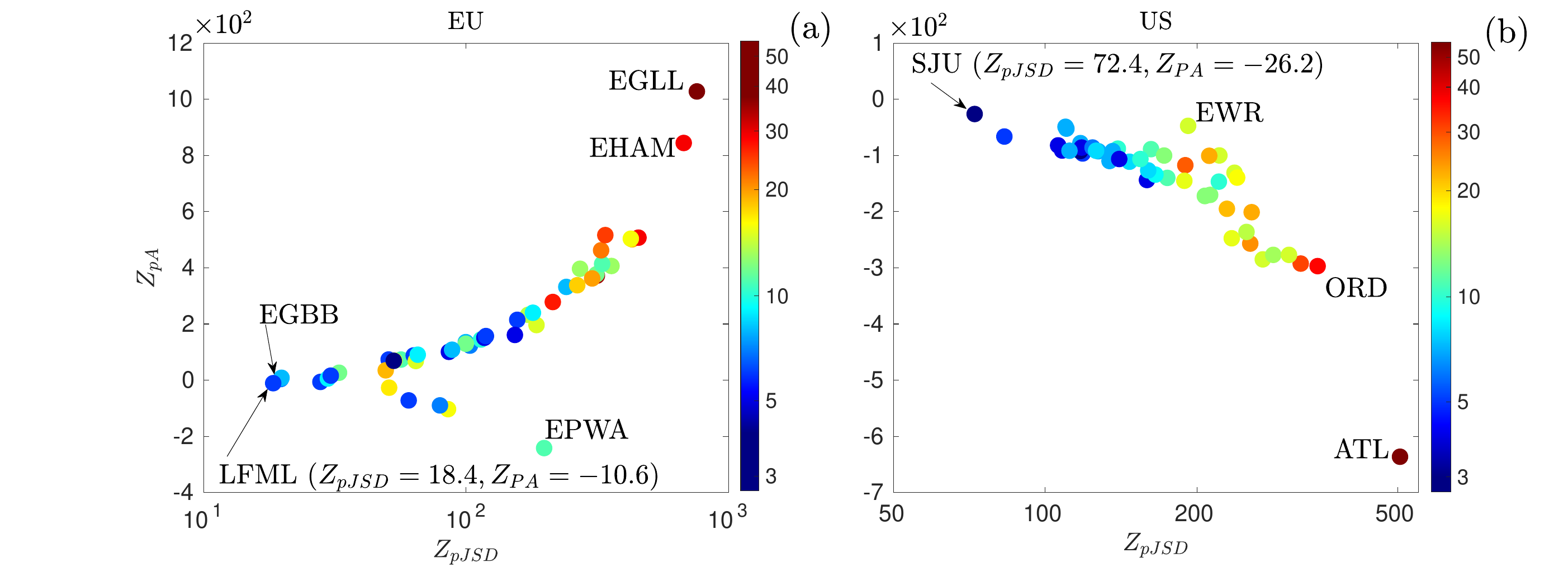}
\caption{\label{fig7} $Z_{pA}$ versus $Z_{pJSD}$ in semi-logarithmic scale for the distribution of arrival delays at the 50 largest airports of (a) Europe and (b) US. The color map indicates the median arrival volume per hour (in logarithmic scale).  }
\end{figure*}

Figure \ref{fig7} reports the asymmetry $Z_{pA}$ as a function of the permutation distance $Z_{pJSD}$, both computed from the ordinal distribution of the walk of the individual arrival delays to their phase-randomized surrogate sequences for (a) EU and (b) US arrival delays. Additionally, the color map indicates the median arrival volume per hour, as a proxy of the size of the airport. Note that similar results are found when using other variables such as the number of arrivals or the number of passengers. Independently of the region, we observe a tendency to draw away from Gaussianity as the arrival volume of the airports increases. To illustrate, in the case of EU, London-Heathrow (EGLL) airport is the farthest away from it and one of the biggest, together with Amsterdam-Schiphol (EHAM)---see Fig. \ref{fig7}(a). Similarly, Atlanta-Hartsfield-Jackson (ATL) and Chicago-O'Hare (ORD) are the furthest from Gaussianity for US, as observed in Fig. \ref{fig7}(b). For the latter region, we found that all individual delay distributions are negative skewed, contrary to the European ones, in which only a few are negative asymmetric, for instance, Varsovia-Chopin (EPWA) airport. These results indicate that, for US, there are more early arrivals than severe delays. Note that no distributions can be categorized as Gaussian, considering that the closest to it are Marsella Provenza (LFML) and Luis Mu\~noz Mar\'in (SJU) airports in EU and US, with $Z_{pJSD}= 18.4$ and 72.4 and $Z_{pA}=-10.6$ and -26.2, respectively---see Fig. \ref{fig7}.   

\begin{figure*}[h!]
\includegraphics[width=\textwidth]{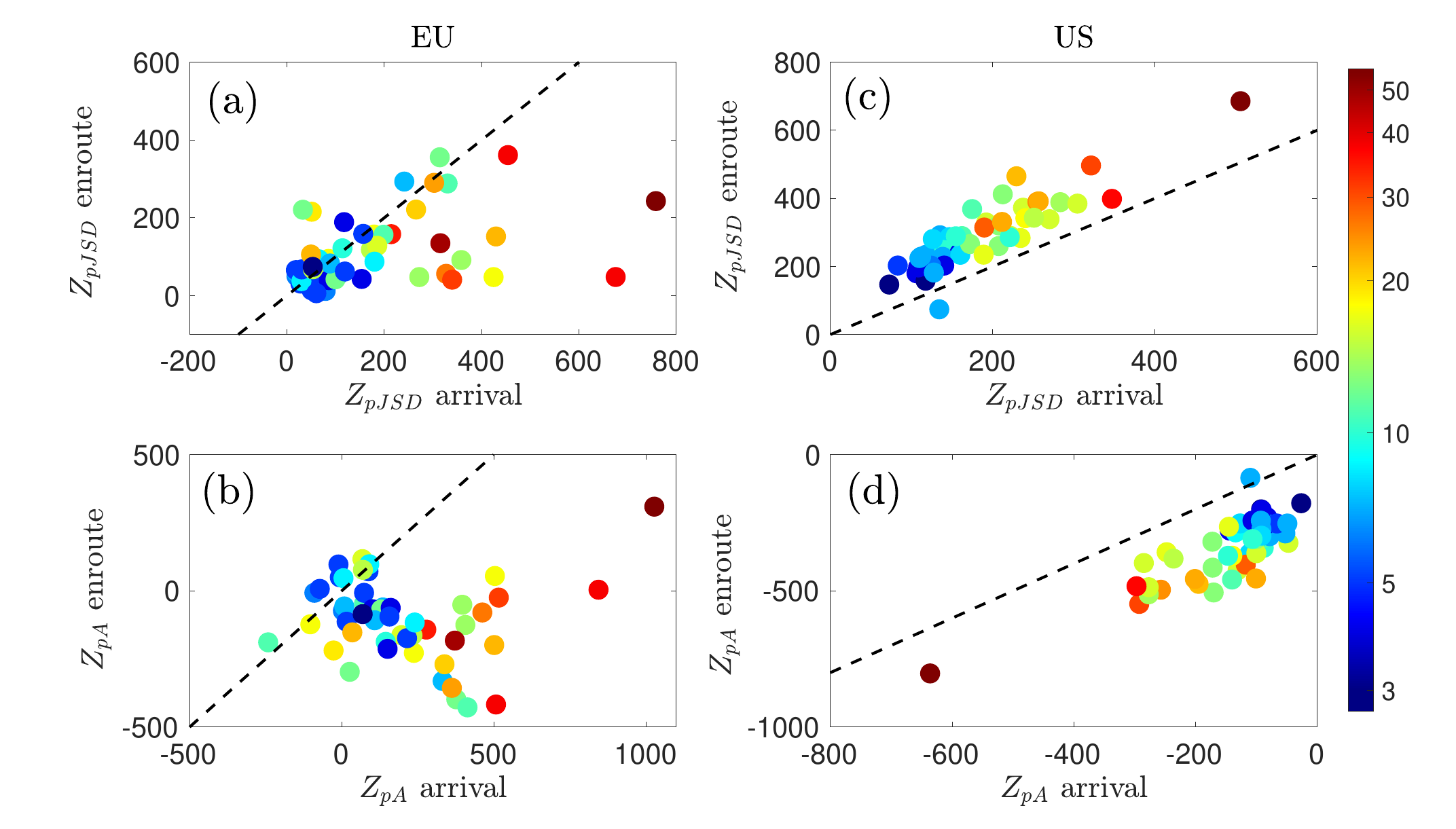}
\caption{\label{fig8} $Z_{pJSD}$ and $Z_{pA}$ for the distributions of en-route versus arrival delays for (a)-(b) EU and (c)-(d) US, respectively. Black dashed line indicates the identity line. The color map indicates the median volume per hour (in logarithmic scale).  }
\end{figure*}

Figure \ref{fig8} shows a scatter plot of the Z-Scores for en-route delays, as a function of those for arrival delays. For EU (left panels), a deviation from Gaussianity similar to the previous one is observed for both distributions. Conversely, in the case of US (right panels), en-route distributions are more distant from the Gaussian assumption. 

In order to understand these results, it is necessary to recall how unexpected events are managed in both systems. In Europe, delays on the ground are prioritised; for instance, when the capacity of the destination airport is reduced, flights are hold on the ground until there is a high certainty about their arrival---this is done through ``slots'', or authorisations to departure within a given time window~\cite{cook2007european}. As a consequence, once a flight has departed, it should not expect delays en-route; and most of the time, the trip ends up being shorted, through directs and other tactical decisions. On the contrary, the US system relies more strongly on delays while en-route, except for some circumstances handled through the Ground Delay Program~\cite{kuhn2013ground}. Note that no approach is inherently better: while the European one avoids en-route delays and their associated environmental impact, the US system is more flexible and can handle disruptions on shorter time scales.

Going back to the results presented in Fig. \ref{fig8}, it can be appreciated that both Z-Scores are highly correlated between en-route and arrival in the US case. In other words, whenever a disruption affects the system, this responds with a similar behaviour in the two conditions, thus in a way aligned with what previously described. The EU system is more complex, with a less clear correlation and more heterogeneous dynamics. On the one hand, $Z_{pA}$ is usually negative: the system tends to favour negative delays while en-route, i.e. aircraft can recover part of their departure delay and do not add additional ones. On the other hand, $Z_{pJSD}$ in the en-route phase are positive and weakly related to the ones in the arrival phase. If a deviation from Gaussianity is interpreted as the appearance of systemic (as opposed to random) disruptions, this result implies that en-route abnormal delays are the result of a disrupted global state. 

\begin{figure*}[h!]
\includegraphics[width=\textwidth]{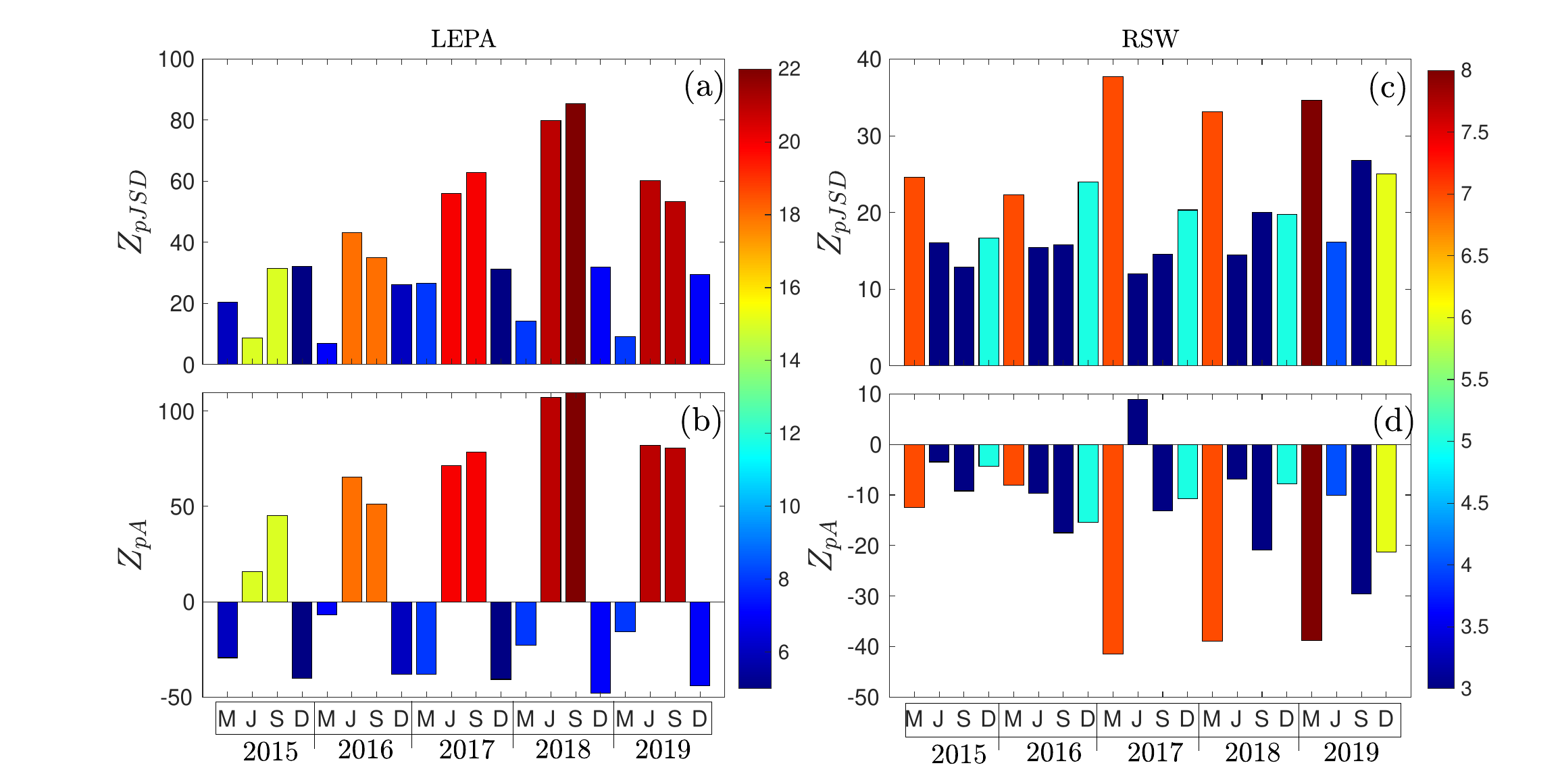}
\caption{\label{fig9} Monthly evolution of the Z-Scores $Z_{pJSD}$ and $Z_{pA}$ for LEPA (a)-(b) and for RSW (c)-(d), respectively. Color maps indicate the median arrival volume per hour.}
\end{figure*}

We further study how deviation from Gaussianity evolves through time. For this, we select two airports for each region of study that present a considerable seasonal variability in traffic and demand. In EU, Mallorca is one of the main tourist destinations in the Mediterranean, attracting over 10 million visitors annually. Its airport Son Sant Joan (LEPA) is the third most important airport in Spain in terms of passenger volume, which makes it a good example to study the monthly change in the distribution of delays. For US, the Southwest Florida International Airport (RSW) is located in a region popular for seasonal visitors (including ``snowbirds''\footnote{``Snowbirds'' is a colloquial term for people (often retirees), who migrate seasonally from colder northern regions to warmer southern areas, such as Florida, during the winter months.}). This airport experiences a steep drop in traffic when the peak winter vacation season ends. Fig. \ref{fig9} reports the results for the monthly analysis of arrival delays at these airports. On one hand, LEPA shows a deviation from Gaussianity perfectly aligned with its size. The permutation asymmetry changes its sign between seasons, being negative for winter months and positive for summer ones---see fig. \ref{fig9}(b)---meaning that there exists a strong correlation between severe delays and the saturation of the airport. On the other hand, even though RSW also shows a correlation between the arrival volume and its distance to its Gaussian reference, its distribution is always negative skewed (except for June 2017), following the general tendency for airports located in the US---compare with fig. \ref{fig7}(b).  

\begin{figure*}[h!]
\includegraphics[width=\textwidth]{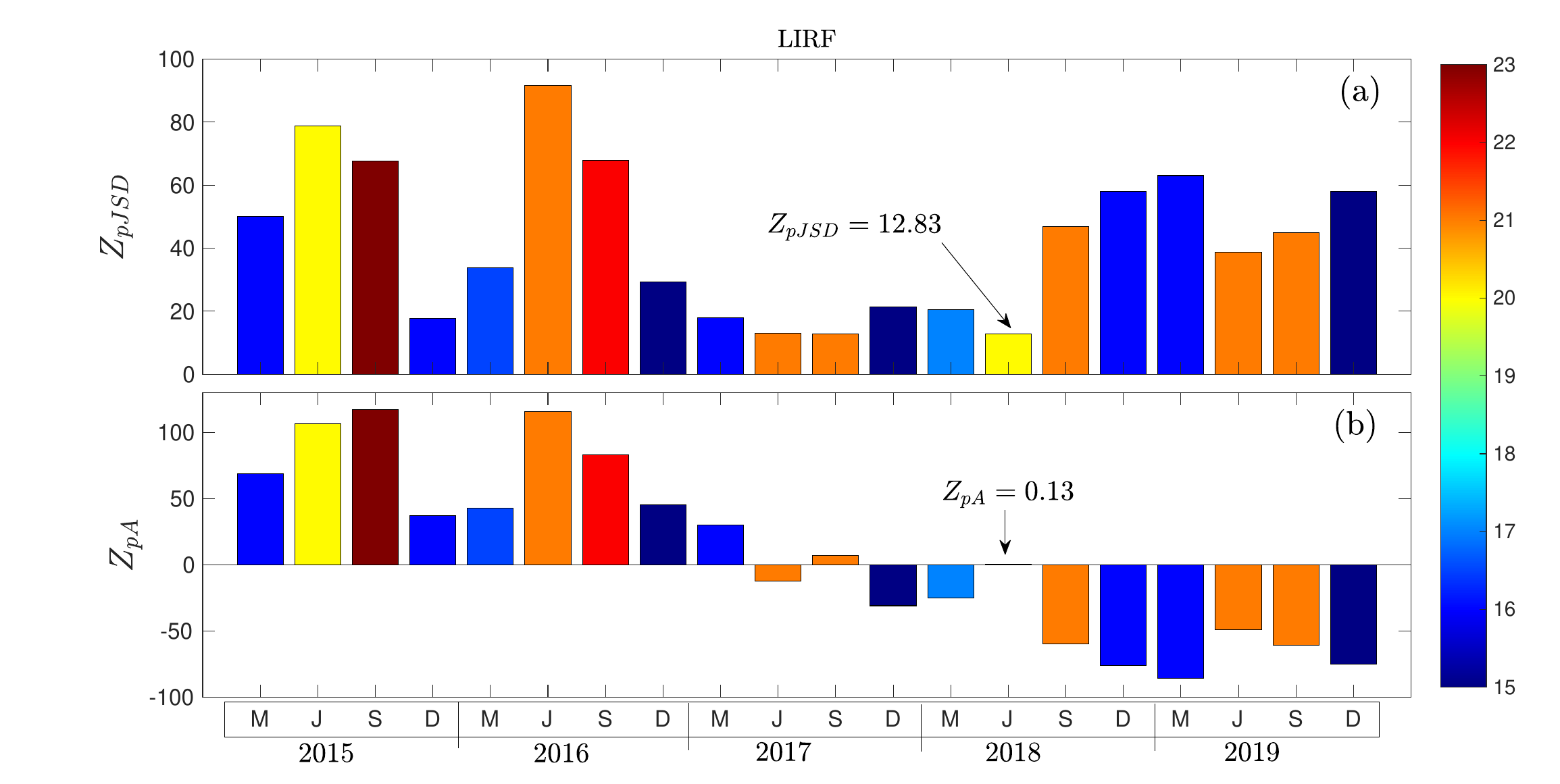}
\caption{\label{fig10} Monthly evolution of the Z-Scores (a) $Z_{pJSD}$ and (b) $Z_{pA}$ for Leonardo da Vinci-Fiumicino Airport (LIRF). Color maps indicate the median arrival volume per hour. }
\end{figure*}

Finally, we study the busiest airport in Italy, serving Rome, the Leonardo da Vinci-Fiumicino Airport (LIRF); this airport has been selected for displaying a very interesting and unusual dynamics. As shown in fig. \ref{fig10}(a), deviations from Gaussianity are quite heterogeneous and do not scale with arrival volume, as the general trend observed for the concatenated data (fig. \ref{fig7}(a)). Even more interesting is the transition found in September of 2017, after which the asymmetry $Z_{pA}$ goes from being positive to negative, implying that the airport went from having more severe delayed arrivals to early ones. Moreover, for June of 2018 we found that the distribution correspond to a Non-Gaussian symmetric one. While it is difficult to provide a complete explanation, several factors may have contributed to this shift in dynamics. Specifically, starting from year 2017, the airport of Rome Fiumicino has experienced a substantial decrease in traffic---mostly due to the problems experienced by the main airline there operating, i.e. Alitalia, currently ITA Airways~\cite{comparison2017}. This, along with more efficient runway operations, a more spread distributions of operations throughout the day and away from peak hours, and the more efficient management of flights as part of the Local Single Sky ImPlementation plan~\cite{pejovic2020relationship}, resulted in a substantial decrease in delays.

\section{Conclusions}

In this study, we have introduced a simple data-driven approach to quantify deviations from Gaussianity in time series. By computing the permutation Jensen-Shannon distance between synthetic data sampled from Stable distributions and their phase-randomized surrogates, we provide a statistical framework that effectively captures the characteristics of heavy tails and skewness. It is noteworthy that, rather than incorporating information about the amplitude distribution when building the ordinal patterns, the present approach demonstrate that quantifying the ordinal dissimilarity between cumulative integration (walk) of the sequences under study and the Gaussian reference, is sufficient to explore non-Gaussian features.

Our findings reveal that flight delay distributions at major European and US airports exhibit significant departures from normality, with US airports predominantly showing negative skewness, indicating a preference for early arrivals over severe delays. In contrast, European airports display more heterogeneous skewness patterns. These results highlight fundamental differences in air traffic management strategies between the two regions. Furthermore, higher traffic volumes correlate with deviations from Gaussianity, particularly in seasonal hubs such as Mallorca and Southwest Florida. This implies that congestion plays a crucial role in shaping the statistical properties of delays. Our study of Rome Fiumicino Airport also revealed a structural transition in its delay dynamics, likely influenced by operational and airline management changes. This result enhances the understanding of non-Gaussian properties in delay distributions, contributing to the development of more accurate forecasting models.

Beyond the characterization of flight delays, our methodology provides a general approach for identifying non-Gaussian signatures in complex systems. It offers a computationally fast, simple and model-free technique that can be applied to other fields where deviations from normality play a critical role, such as financial time series~\cite{rak2018quantitative} and physiologial data~\cite{peng1993long}.

\section*{Funding}
This project has received funding from the European Research Council (ERC) under the European Union's Horizon 2020 research and innovation programme (grant agreement No 851255). This work was partially supported by the Mar\'ia de Maeztu project CEX2021-001164-M funded by the  MICIU/AEI/10.13039/501100011033.

\bibliographystyle{elsarticle-num}
\bibliography{main.bib}

\end{document}